\documentclass[lettersize,journal]{IEEEtran}
\usepackage{amsmath,amsfonts}
\usepackage{algorithmic}
\usepackage{algorithm}
\usepackage{array}
\usepackage[caption=false,font=normalsize,labelfont=sf,textfont=sf]{subfig}
\usepackage{textcomp}
\usepackage{stfloats}
\usepackage{url}
\usepackage{verbatim}
\usepackage{graphicx}
\usepackage{color}
\usepackage{cite}
\begin{document}
\title{Deep Learning Based Joint Beamforming Design\\ in IRS-Assisted Secure Communications}
\author{Chi Zhang, Yiliang Liu,\textit{ Member, IEEE}, and Hsiao-Hwa Chen,\textit{ Fellow, IEEE}
\thanks{C. Zhang (email: {\tt maye1998@163.com}) is with the School of Electronics and Information Engineering, Harbin Institute of Technology, China. Y. Liu (email: {\tt liuyiliang@xjtu.edu.cn}) is with the School of Cyber Science and Engineering, Xi'an Jiaotong University, China. Hsiao-Hwa Chen (email: {\tt hshwchen@mail.ncku.edu.tw}) is with the Department of Engineering Science, National Cheng Kung University, Taiwan.}}

\maketitle

\begin{abstract}
In this article, physical layer security (PLS) in an intelligent reflecting surface (IRS) assisted multiple-input multiple-output multiple antenna eavesdropper (MIMOME) system is studied. In particular, we consider a practical scenario without instantaneous channel state information (CSI) of the eavesdropper and assume that the eavesdropping channel is a Rayleigh channel. To reduce the complexity of currently available IRS-assisted PLS schemes, we propose a low-complexity deep learning (DL) based approach to design transmitter beamforming and IRS jointly, where the precoding vector and phase shift matrix are designed to minimize the secrecy outage probability. Simulation results demonstrate that the proposed DL-based approach can achieve a similar performance of that with conventional alternating optimization (AO) algorithms for a significant reduction in the computational complexity.
\end{abstract}

\section{INTRODUCTION}
Intelligent reflecting surface (IRS) is a 2-D electromagnetic metasurface composed of a large number of low-cost passive elements that can change wireless propagation environment to improve the performance of wireless communication systems \cite{wu2019towards}. Physical layer security (PLS) is a type of information security technique that utilizes the randomness of wireless channels to achieve  confidential communications \cite{liu2016physical}. As IRS can improve the quality of legitimate channels while mitigating that of eavesdropped channels, the two techniques combined have a complementary nature of advantages \cite{almohamad2020smart}.

In \cite{shen2019secrecy} and \cite{cui2019secure}, the authors investigated PLS of IRS-assisted single-input single-output (SISO) and multiple-input single-output (MISO) communication systems to maximize secrecy rate of the system by alternating optimization (AO) algorithms. In \cite{xu2019resource}, the authors extended it to a multiple-user case that transmits artificial noise (AN) to enhance the security performance. Moreover, \cite{xu2022intelligent} focused on a novel idea of IRS-assisted PLS, in which IRS was deployed to modulate received signal from a transmitter as AN signals. However, the above works were based largely on the assumption that eavesdropper's instantaneous channel state information (CSI) is known. The authors of \cite{wang2020intelligent} investigated the PLS by jointly optimizing precoding, IRS phase shift matrix, and AN when the eavesdropper's CSI was unknown, but there was no appropriate metric to evaluate the performance of PLS. The work in \cite{feng2020large} used a large number of random samples instead of eavesdropper's statistical CSI, but the proposed optimization algorithm needs to take into account all random samples, which consumes a lot of computational resources. In \cite{liu2022min}, the authors investigated the PLS based on eavesdropper's statistical CSI and derived a secrecy outage probability expression as a security metric. However, the complexity of the proposed AO algorithm for the optimization problem of minimizing secrecy outage probability is very high.

To address the CSI issue, we consider an IRS-assisted PLS scenario based on the eavesdropper's statistical CSI. Different from~\cite{liu2022min}, we think the large-scale fading to be non-negligible, so we consider the large-scale fading in the calculation of the secrecy outage probability. Motivated by the machine learning-based research works \cite{liu2020deep, song2020unsupervised, elbir2021federated}, we propose a low-complexity deep learning (DL) based approach to design transmitter beamforming and phase shift matrix to minimize the secrecy outage probability, which works based on unsupervised learning and does not need dataset labels.

\section{SYSTEM MODEL}
\begin{figure}[htb]
\centering
\includegraphics[width = 0.4\textwidth]{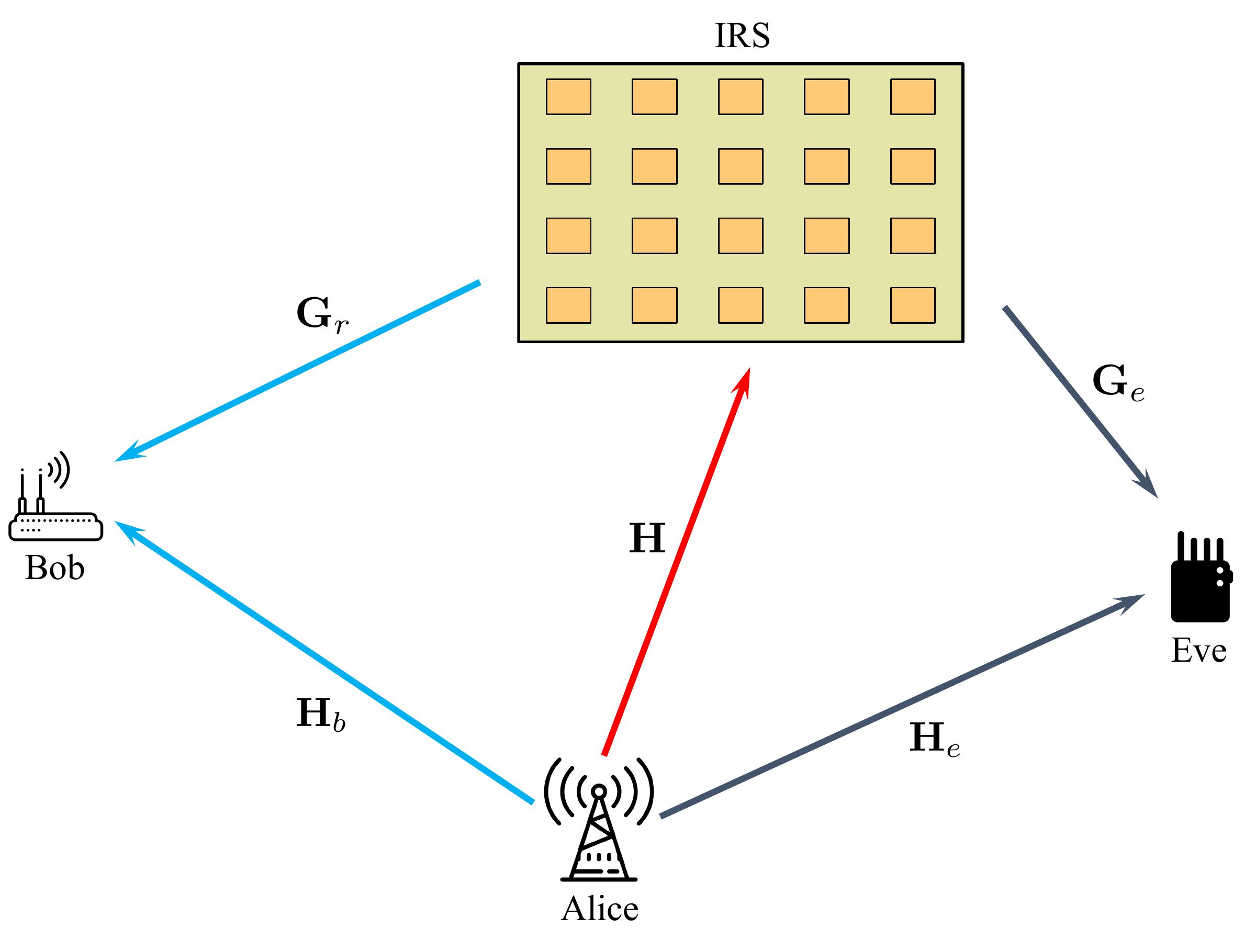}
\caption{An IRS-assisted MIMOME system.}
\label{fig:1}
\end{figure}

Let us consider an IRS-assisted multiple-input multiple-output multipleantenna eavesdropper (MIMOME) system, as shown in Fig. \ref{fig:1}, where a transmitter (Alice) equipped with $N_t$ antennas serves an $N_r$-antenna legitimate user (Bob), assisted by an IRS with $N_s$ reflecting elements. Meanwhile, there is an eavesdropper (Eve) with $N_e$ antennas that receives signals from Alice and IRS. The legitimate channels from Alice to Bob, Alice to IRS, and IRS to Bob are denoted by $\mathbf{H}_b\in \mathbb{C}^{N_r\times N_{t}}$, $\mathbf{H}\in\mathbb{C}^{N_s\times N_{t}}$, and $\mathbf{G}_r\in\mathbb{C}^{N_r\times N_{s}}$, respectively. The wiretap channels from Alice to Eve and IRS to Eve are denoted by $\mathbf{H}_e \in \mathbb{C}^{N_e \times N_{t}}$ and $\mathbf{G}_e \in \mathbb{C}^{N_e \times N_{s}}$, respectively. Assume that all channels follow the quasi-static flat-fading channel model, and the legitimate channels $\mathbf{H}_b$, $\mathbf{G}_r$, and $\mathbf{H}$ can be perfectly estimated at Alice \cite{wei2021channel}, while the instantaneous CSIs of wiretap channels $\mathbf{H}_e$ and $\mathbf{G}_e$ are unknown to Alice. In addition, we further assume that the wiretap channels obey Rayleigh fading without loss of generality.

Let $s$ be the transmitted single-stream confidential signal following $\mathbb{E}\left[|s|^{2}\right]=1$. Then, the received signals at Bob and Eve can be expressed as
\begin{align}
&\mathbf{y}=\left(\mathbf{H}_{b}+\mathbf{G}_{r} {\mathbf\Theta \mathbf{H}}\right) \mathbf{w}s+\mathbf{n}, \label{eq:1}\\
&\mathbf{y}_{e}=\left(\mathbf{H}_{e}+\mathbf{G}_{e} {\mathbf\Theta \mathbf{H}}\right) \mathbf{w}s+\mathbf{n}_{e}, \label{eq:2}
\end{align}
where $\mathbf\Theta  = {\mathrm{diag}}\left( {{e^{j{\theta _1}}},{e^{j{\theta _2}}}, \cdots ,{e^{j{\theta _{N_s}}}}} \right)$ denotes a phase shift matrix and $\theta_n$ is the phase shift introduced by the $n$th element of IRS. $\mathbf{w} \in \mathbb{C}^{N_t \times 1}$ is the precoding vector and satisfies $\| \mathbf{w} \|^2 = P_t$, where $P_t$ is the transmit power of Alice. $\mathbf{n} \sim \mathcal{C} \mathcal{N}\left(\mathbf{0}, \sigma^2\mathbf{I}_{N_t}\right)$ and $\mathbf{n}_e \sim \mathcal{C} \mathcal{N}\left(\mathbf{0}, \sigma_e^2\mathbf{I}_{N_e}\right)$ are additive white Gaussian noise (AWGN) at Bob and Eve, respectively.

To maximize the signal-noise-ratio (SNR) at Bob, the corresponding received signal vector $\mathbf{w}_{r}=\left[\left(\mathbf{H}_{b}+\mathbf{G}_{r} {\mathbf\Theta \mathbf H}\right) \mathbf{w}\right]^{{H}} /  \|\left(\mathbf{H}_{b}+\mathbf{G}_{r} {\mathbf\Theta \mathbf H}\right) \mathbf{w}\|$ is given with the maximum ratio combining (MRC) strategy \cite{liu2016physical}. We consider a worst-case scenario that Eve has perfect knowledge about $\mathbf{H}_{e}$, $\mathbf{G}_{e}$, $\mathbf{H}$, $\mathbf{G}_r$, $\mathbf\Theta$, and $\mathbf{w}$ within the coherence time. Then, Eve can also maximize the SNR using the MRC strategy. Accordingly, the channel capacities at Bob and Eve are given by
\begin{align}
C_{m} &=\log _{2}\Big[1+\frac{P_t}{\sigma^{2}}\|\left(\mathbf{H}_{b}+\mathbf{G}_r {\mathbf\Theta} \mathbf{H}\right) \mathbf{b}\|^{2}\Big], \label{eq:3}\\
C_{w} &=\log _{2}\Big[1+\frac{P_t}{\sigma_{e}^{2}}\|\left(\mathbf{H}_{e}+\mathbf{G}_{e} {\mathbf\Theta} \mathbf{H}\right) \mathbf{b}\|^{2}\Big], \label{eq:4}
\end{align}
where $C_m$ and $C_w$ are the main channel capacity and wiretap channel capacity, respectively, and $\mathbf{b}=\mathbf{w}/\sqrt{P_t}$ is the normalized precoding vector. 

However, due to the wiretap channels $\mathbf{H}_e$ and $\mathbf{G}_e$ which are unknown, it is not possible to calculate the secrecy capacity $C_s$, where $C_s=[C_m-C_w]^+$. Therefore, we choose secrecy outage probability as a security metric for PLS, which is defined as the probability that the secrecy capacity $C_s$ is smaller than the target PLS coding rate $R_s$. The secrecy outage probability is defined as
\begin{subequations}
\begin{align}
P_{\text {out }}\left(R_{s}\right) &=P\left(C_{s} \leq R_{s} ~\mid\text{Transmission}\right) \label{eq:5a} \\ 
&=P\left(\|\left(\mathbf{H}_{e}+\mathbf{G}_{e} {\mathbf\Theta \mathbf H}\right) \mathbf{b}\|^{2} \geq  \phi\right), \label{eq:5b}
\end{align}
\end{subequations}
where $\phi=\sigma_{e}^{2}\left(2^{C_{m}-R_{s}}-1\right) / P_t$.

Assume that wiretap channels $\mathbf{H}_e$ and $\mathbf{G}_e$ obey Rayleigh fading in this work, i.e., $\mathbf{H}_{e}=\beta_{d} \mathbf{Z}$ and $\mathbf{G}_{e}=\beta_{r} \mathbf{C}$, where $\mathbf{Z} \sim \mathcal{C} \mathcal{N}_{N_{e}, N_{t}}\left(\mathbf{0}, \mathbf{I}_{N_{e}} \otimes \mathbf{I}_{N_{t}}\right)$ and $\mathbf{C} \sim \mathcal{C} \mathcal{N}_{N_{e}, N_{s}}\left(\mathbf{0}, \mathbf{I}_{N_{e}} \otimes \mathbf{I}_{N_{s}}\right)$. Note that $\otimes$, $\beta_d$, and $\beta_r$ denote the Kronecker products, large-scale fading factors of $\mathbf{H}_e$ and $\mathbf{G}_e$, respectively. 

To derive an expression of the secrecy outage probability, we introduce an auxiliary random variable $X$ as follows.
\begin{subequations}
\begin{align}
X &= \|\left(\mathbf{H}_{e}+\mathbf{G}_{e} {\mathbf\Theta \mathbf H}\right) \mathbf{b}\|^2 \label{eq:6a}\\ 
&=\|\beta_d \mathbf{Z}\mathbf{b}+\beta_r \mathbf{C} {\mathbf\Theta \mathbf H}\mathbf{b}\|^2. \label{eq:6b}
\end{align}
\end{subequations}
The secrecy outage probability can be given by
\begin{equation}
P_{\mathrm{out}}\left(R_{s}\right)=1-F_X \left( \phi \right),
\label{eq:6}
\end{equation}
where $F_X(\cdot)$ is the cumulative distribution function (CDF) of $X$. With the help of a Gamma distribution \cite{liu2022min,van2020coverage}, we can deduce an approximate expression of secrecy outage probability as\footnote{For the detailed derivation of the Eq. (\ref{eq:8}), please refer to \url{https://github.com/MayeZhang/DL-IRSBF-PLS.}} 
\begin{equation}
\label{eq:8}
P_{\text {out }}\left(R_{s}\right) =\frac{1}{\Gamma\left(N_{e}\right)} \Gamma\left(N_{e}, \frac{\phi}{\beta_d^{2}+\beta_r^2\|\mathbf{\Theta H b}\|^{2}}\right),
\end{equation}
where $\Gamma(m)=\int_{0}^{\infty} t^{m-1} e^ {-t} ~\mathrm{d} t$ is the Gamma function, and $\Gamma(m, n)=\int_{n}^{\infty} t^{m-1} e^ {-t} ~\mathrm{d} t$ is the upper incomplete Gamma function.  

In this work, we aim to design joint beamforming by optimizing $\mathbf{\Theta}$ and $\mathbf{b}$ at Alice and IRS to minimize the secrecy outage probability. Moreover, the secrecy outage probability $P_{\text {out }}$ decreases as $\phi /\left( \beta_d^{2}+\beta_r^2\|\mathbf{\Theta H b}\|^{2}\right)$ increases \cite{liu2022min}. Thus, the optimization problem can be formulated as
\begin{subequations}
\label{eq:9}
\begin{align}
\max \limits_{\mathbf{\Theta}, \mathbf{b}} ~~& \frac{\phi}{\beta_d^{2}+\beta_r^2\|\mathbf{\Theta H b}\|^{2}}, \label{eq:9a}\\
\mathrm { s.t. } ~~&\mathbf{b}^H \mathbf{b}=1, \\
& |e^{j\theta_n} |=1, \forall n=1, \cdots, N_s. \label{eq:9c}
\end{align}
\end{subequations}
It is worth noting that the optimal global solution of the optimization problem with a non-convex unit modulus constraint Eq. (\ref{eq:9c}) is not available. Therefore, \cite{liu2022min} proposed a conventional AO algorithm to solve it, but the computational complexity is very high. In the next section, we will propose a DL-based approach that has a much lower computational complexity.

\section{DL-BASED JOINT BEAMFORMING DESIGN}
\begin{figure}[htb]
\centering
\includegraphics[width = 0.45\textwidth]{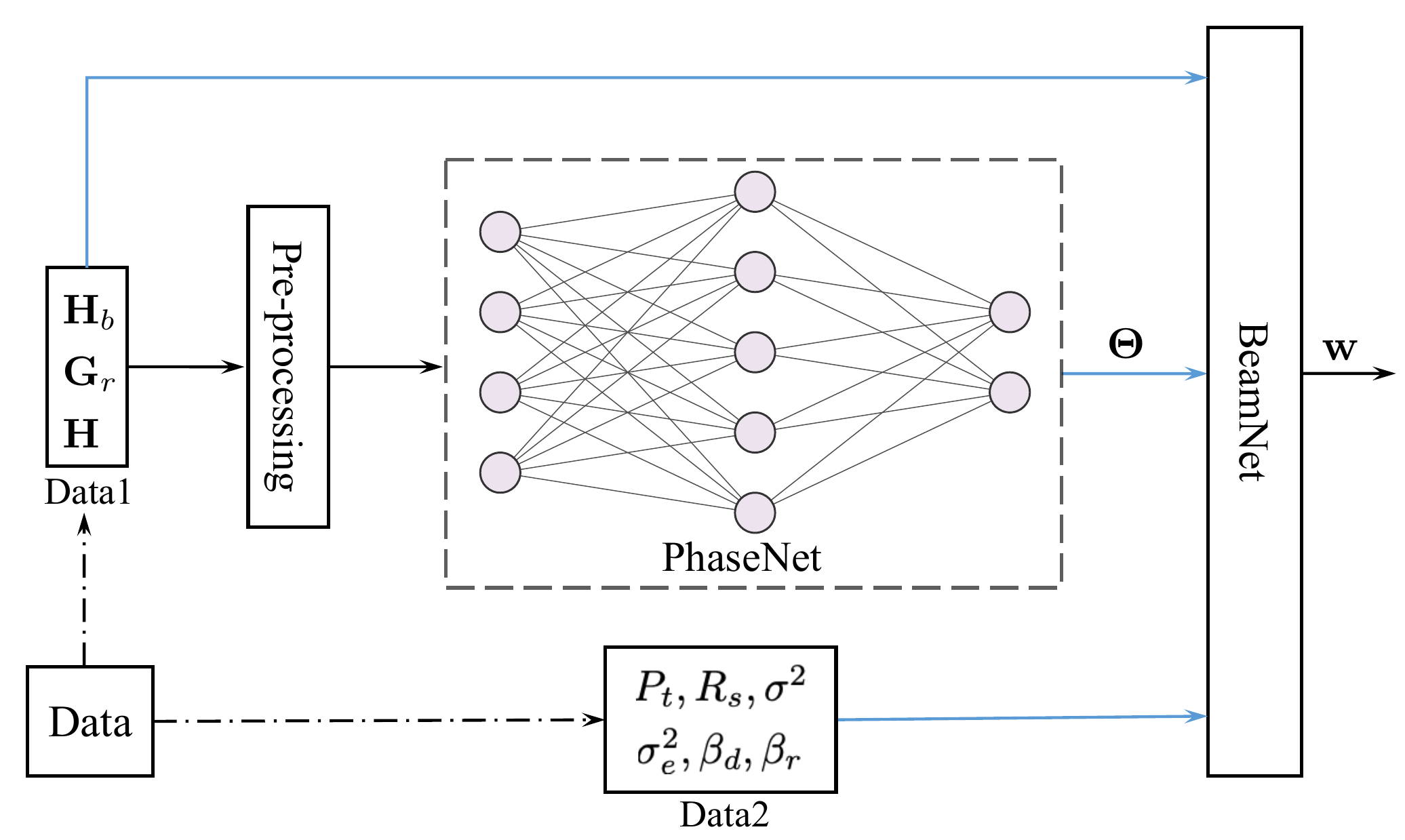}
\caption{Proposed JBNet architecture for joint beamforming design, where data is composed of Data1 and Data2.}
\label{fig:2}
\end{figure}

In this section, we propose a low complexity DL-based joint beamforming network (JBFNet) to solve the problem (\ref{eq:9}). The detailed JBFNet architecture is shown in Fig. \ref{fig:2}, which is divided into two parts. One part is named ``PhaseNet'' for predicting the phase shift matrix $\mathbf \Theta$ of IRS, and the other part is named "BeamNet" for predicting the precoding vector $\mathbf w$. In the following subsections, we discuss the issues on the design of JBFNet in detail.

\subsection{PhaseNet}
\begin{figure}[htb]
\centering
\includegraphics[width = 0.5\textwidth]{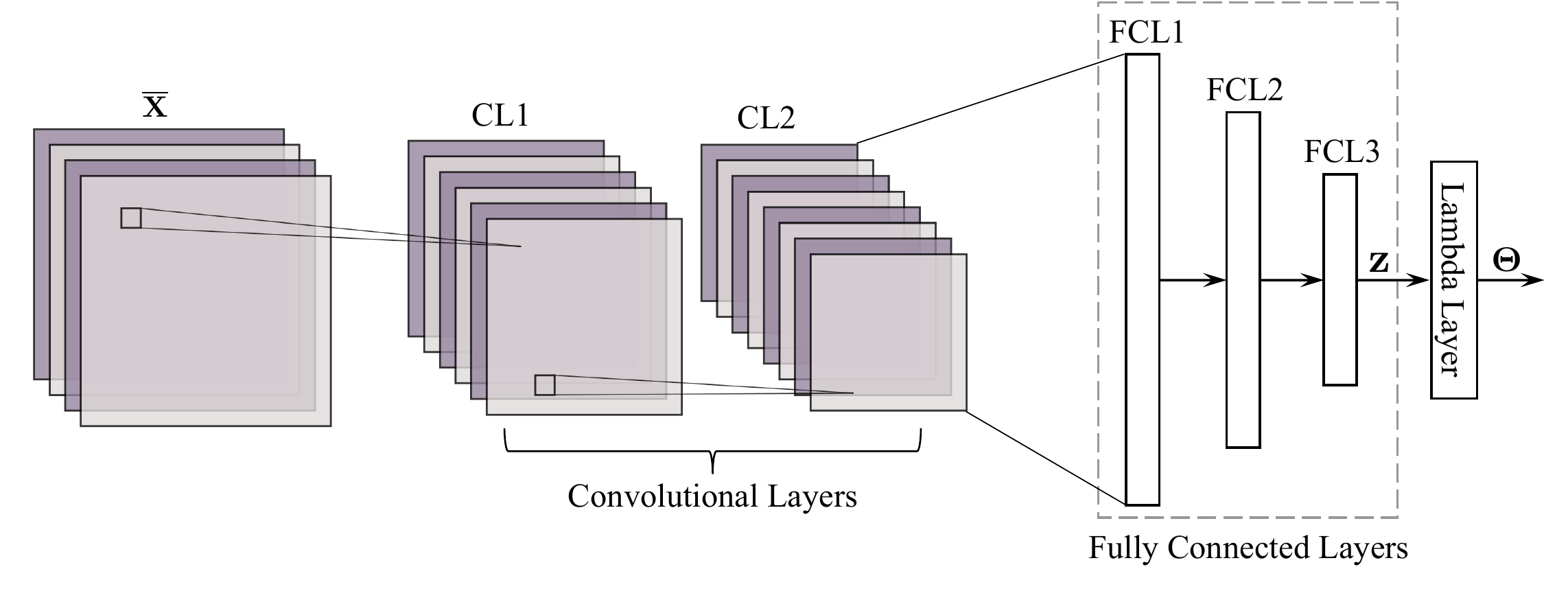}
\caption{PhaseNet framework for $\mathbf{\Theta}$ based on CNN.}
\label{fig:3}
\end{figure}

PhaseNet adopts a multi-layer convolutional neural network (CNN) structure as shown in Fig. \ref{fig:3}, which is commonly used in image processing problems and has been proven to be very effective in communication problems like beamforming design and channel estimation \cite{liu2020deep, song2020unsupervised}. In addition, compared with the commonly used fully connected network (FCN), CNN has a stronger feature extraction capability of 2-D data.

According to \cite{gao2020unsupervised}, the phase shift matrix $\mathbf{\Theta}$ can be obtained by training the channel data, such that we choose the legitimate channels $\mathbf{H}_b$, $\mathbf{G}_r$, and $\mathbf{H}$ as the input to PhaseNet, which are 2-D matrices. In order to improve JBNet's capability in feature extraction, the input $\text{Data1}=\{\mathbf{H}_b, \mathbf{G}_r,  \mathbf{H} \}$ needs to be pre-processed. Define $\mathbf{F}_n =\mathbf{g}_{r,n}\mathbf{h}_n^H$, where $\mathbf{g}_{r,n}$ is the $n$th column of $\mathbf{G}_r$, and $\mathbf{h}_n^H$ is the $n$th row of $\mathbf{H}$. Specifically, $\mathbf{F}_n$ is also the cascaded channel corresponding to the $n$th element of IRS. Then the channel in Eq. (\ref{eq:1}) can be rewritten as
\begin{equation}
\label{eq:10}
\mathbf{H}_b + \mathbf{G}_r\mathbf{\Theta H}=\mathbf{H}_b+\sum_{n=1}^{N_s} e^{j\theta_n}\mathbf{F}_n.
\end{equation}
Specifically, the original data $\mathbf{X} \in \mathbb{C}^{N_r \times N_t \times (N_s + 1)}$ can be viewed as a $N_r \times N_t$ image with $N_s+1$ channels, i.e., $\mathbf{X}=\left[\mathbf{H}_b; \mathbf{F}_1; \mathbf{F}_2;\cdots;\mathbf{F}_{N_s} \right]$. The input data of the neural network are required to take real values, but the original data $\mathbf X$ consisting of CSI are the complex values. To deal with this issue, the real part, imaginary part, and absolute value of each element of $\mathbf{X}$ are extracted to form a new 3-D matrix, i.e., $\overline {\mathbf{X}}=\{ \Re{\left(\mathbf{X}\right)}, \Im{\left(\mathbf{X}\right)}, \text{Abs}{\left(\mathbf{X}\right)}\}$. Note that $\Re{\left(\mathbf{X}\right)}$, $\Im{\left(\mathbf{X}\right)}$, and $\text{Abs}{\left(\mathbf{X}\right)}$ are combined according to the CNN channel dimension by default, i.e., {$\overline{\mathbf{X}} \in \mathbb{R}^{N_r \times N_t \times 3(N_s + 1)} $. In addition, the dimension can be flexibly selected for combination according to the actual parameter settings. For instance, when the number of receiving antennas $N_r=2$, resulting in a small ``image'' that is not conducive to the feature extraction, the combination can be done by the first dimension, i.e., {$\overline{\mathbf X} \in \mathbb{R}^{3N_r \times N_t \times (N_s + 1)}$}.}

As shown in Fig. \ref{fig:3}, the preprocessed data $\overline{\mathbf X}$ passes through two convolutional layers (CLs) and three fully connected layers (FCLs), and the results are finally outputted by the Lambda layer. We design CL1 with 256 filters of size $2\times2$, CL2 with 512 filters of size $2\times 2$, FCL1 and FCL2 with 64$N_s$ and 16$N_s$ neurons, respectively. In particular, the number of neurons of FCL3 is set to $N_s$, which is the same as the number of IRS reflecting elements. In order to prevent overfitting of the neural network, a batch normalization (BN) layer is inserted between every two layers, and we take rectified linear unit (ReLU) as the activation function except for FCL3. Specifically, FCL3 adopts ``Sigmoid'' as the activation function, so that the value of the output $\mathbf z$ can be compressed between $(0, 1)$. Then, we can get the IRS phase shift matrix as $\mathbf \Theta = \mathrm{diag}\left[\exp(j \cdot 2\pi \mathbf{z}) \right]$ through Lambda layer.

\subsection{BeamNet}
For any given phase shift matrix $\mathbf \Theta$, the objective function Eq. (\ref{eq:9a}) is transformed as
\begin{equation}
\label{eq:11}
\frac{\phi}{\beta_d^{2}+\beta_r^2\|\mathbf{\Theta H b}\|^{2}}=c\left( \frac{\mathbf{b}^H \left(t\mathbf{I}_{N_t} + \mathbf{Q}_1\right) \mathbf{b}}{\mathbf{b}^H \left(\beta_d^2 \mathbf{I}_{N_t} + \mathbf{Q}_2\right) \mathbf{b}}  \right),
\end{equation}
where $c=\sigma_e^2 / (2^{R_s}\sigma^2)$, $t=\sigma^2(1-2^{R_s})/P_t$, $\mathbf{Q}_1=\left( \mathbf{H}_{b}+\mathbf{G}_r {\mathbf\Theta} \mathbf{H} \right)^H \left( \mathbf{H}_{b}+\mathbf{G}_r {\mathbf\Theta} \mathbf{H} \right)$, and $\mathbf{Q}_2=\beta_r^2\mathbf{H}^H\mathbf{H}$. According to the generalized Rayleigh quotient, the optimal $\mathbf{b}$ is given by 
\begin{equation}
\label{eq:12}
\mathbf{b}^{*}=\text { eigvec }_{\lambda_{\max }}\left[\left(\beta_d^{2} \mathbf{I}_{N_{t}}+\mathbf{Q}_{2}\right)^{-1}\left(t \mathbf{I}_{N_{t}}+\mathbf{Q}_{1}\right)\right],
\end{equation}
where $\text {eigvec}_{\lambda_{\max }}(\mathbf{X})$ is the eigenvector corresponding to the largest eigenvalue of matrix $\mathbf{X}$, and $\lambda_{\max }$ is the largest eigenvalue of $\mathbf{X}$. Thus, the optimal beamforming vector can be given by $\mathbf{w}^{*}=\sqrt{P_t}\mathbf{b}^{*}$.

In summary, BeamNet yields the optimal $\mathbf{w}^{*}$ by computing Eq. (\ref{eq:12}) on the basis of the predicted $\mathbf \Theta$ of PhaseNet, Data1, and $\text{Data2}=\{ P_t, R_s, \sigma^2, \sigma_e^2, \beta_d, \beta_r\}$.

\subsection{Loss Function}
Different from traditional supervised learning, JBNet adopts unsupervised learning, which does not need additional labels of $\mathbf{\Theta}$ and $\mathbf{b}$. In our design, the loss function is related directly to the objective function Eq. (\ref{eq:9a})
\begin{equation}
\label{eq:13}
\mathrm{Loss}= -\frac{1}{K}\sum_{k=1}^{K}\frac{\phi_k}{\left(\beta_{d,k}^{2}+\beta_{r,k}^2\|\mathbf{\Theta}_k \mathbf{H}_k \mathbf{b}_k\|^{2} \right)},
\end{equation}
where $K$ is the number of the samples in each batch of the training set. Note that the neural network is trained in the direction of minimizing the loss function, which exactly corresponds to an increasing $\phi /\left( \beta_d^{2}+\beta_r^2\|\mathbf{\Theta H b}\|^{2}\right)$ in Eq.~(\ref{eq:9a}).

\subsection{Network Training}
According to PhaseNet and BeamNet, the entire input data of JBNet is composed of Data1 and Data2, i.e., $\text{Data}=\{ \mathbf{H_b}, \mathbf{G_r}, \mathbf{H}, P_t, R_s, \sigma^2, \sigma_e^2, \beta_d, \beta_r\}$. To train JBFNet adequately, we generated $8\times 10^5$ and $2\times 10^5$ data samples for training and validation randomly, respectively. Furthermore, we set the maximum training epochs as 2000 and the batch size as 1000. The optimizer is set to perform adaptive moment estimation (Adam) with an initial learning rate of 0.01. In order to accelerate the convergence and prevent overfitting, the learning rate decays by a factor of 0.3 when the loss on the validation does not decrease for 15 consecutive epochs, and an early stop with patience 20 is applied. All training processes are implemented with Python 3.8, Cuda 11.1, and PyTorch 1.9.0 on a PC equipped with a GeForce GTX 1080 Ti GPU.

\section{SIMULATION RESULTS}
In this section, we give the simulation results of the proposed DL-based method. The simulations adopted the parameters of $N_t=4$, $N_r=2$, $N_e=2$, the noise variances $\sigma^2= \sigma_e^2$ are normalized to one, and $P_t$ is defined in dB with respect to $\sigma^2$. The legitimate channels $\mathbf{H}_b$, $\mathbf{H}$, and $\mathbf{G}_r$ are assumed to be independent Rayleigh fading, and the large-scale fading factors of the wiretap channels $\beta_d$ and $\beta_r$ were generated randomly within $(0, 1)$. All simulation results are averaged over 1000 channel realizations, and we compare the performance of the proposed DL-based approach with the following schemes:
\begin{itemize}
\item \textbf{Without IRS}: The precoding vector $\mathbf w$ is given by the maximum ratio transmission (MRT) strategy, i.e., $\mathbf{w}=\sqrt{P_t} \text {eigvec}_{\lambda_{\max }}\left( \mathbf{H}_b^H \mathbf{H}_b \right)$. According to Eq. (\ref{eq:8}), the secrecy outage probability is calculated as $P_{\text {out }}\left(R_{s}\right)=\Gamma\left( N_e, \phi/\beta_d^2 \right)/\Gamma(N_e)$, where $\phi=\sigma_{e}^{2}\left(2^{C_{m}-R_{s}}-1\right) / P_t$ and $C_m=\log_2\left(1+P_t\lambda_{\max}/\sigma^2 \right)$. 
\item \textbf{Random phase}: The phase shift $\theta_n$ of each element is generated randomly at $\left[0, 2\pi \right]$, and then $\mathbf{w}$ is calculated by the random $\mathbf{\Theta}$ and Eq. (\ref{eq:12}).
\item \textbf{AO algorithm}: The optimization problem Eq. (\ref{eq:9}) is solved by AO-SDR algorithm and AO-Man algorithm, which are proposed in \cite{liu2022min}. 
\end{itemize}

\begin{figure}[htb]
\centering
\includegraphics[width = 0.45\textwidth]{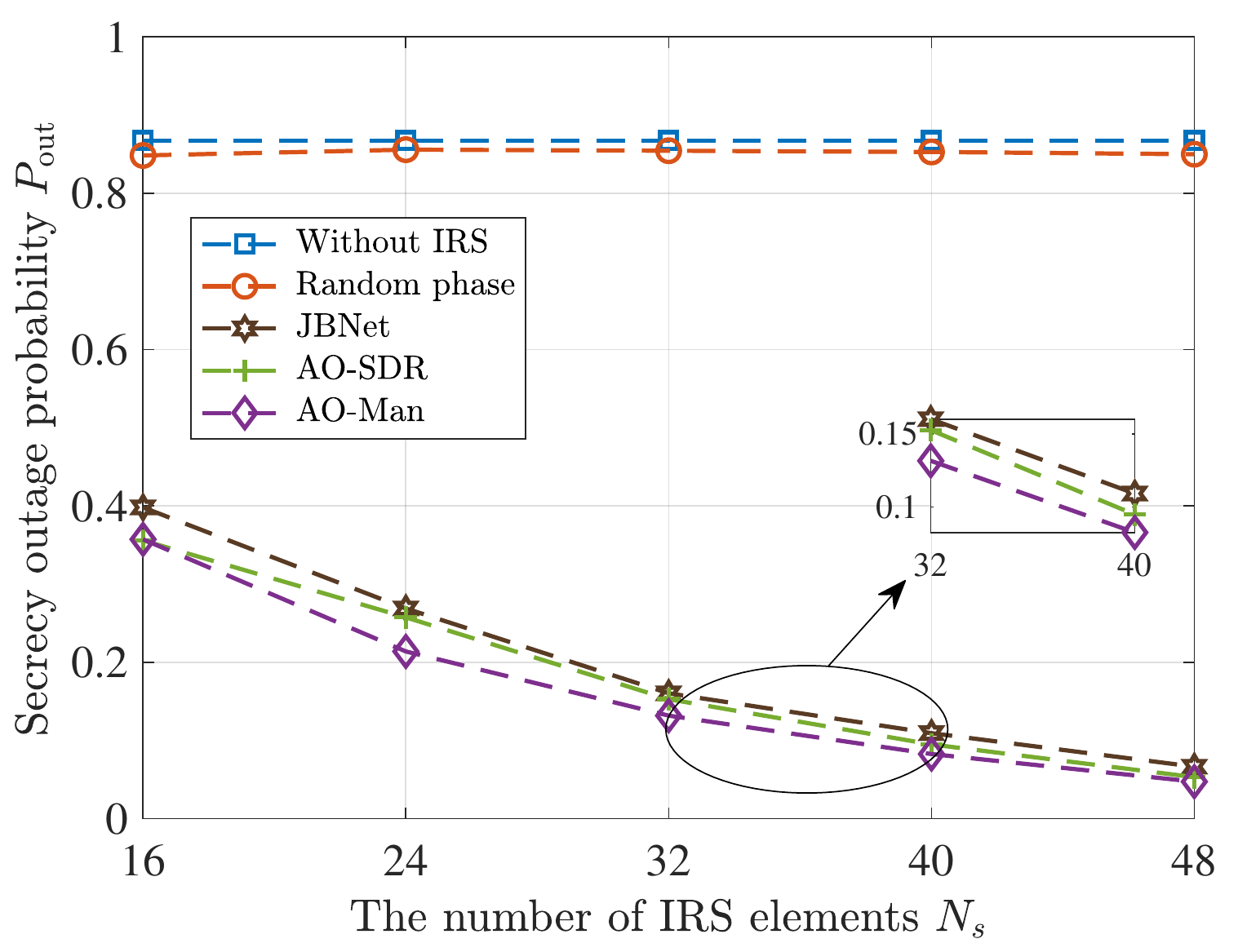}
\caption{Secrecy outage probability in terms of $N_s$ when $\text{SNR}=10~\text{dB}$ and $R_s=3.5 ~\text{bit/s/Hz}$.}
\label{fig:4}
\end{figure}

Fig. \ref{fig:4} shows the secrecy outage probability in terms of different numbers of IRS elements $N_s$. As we can observe, the proposed JBNet and two AO algorithms (AO-SDR and AO-Man) reduce the secrecy outage probability as $N_s$ increases, and there is a small gap between JBNet and the two AO algorithms. Moreover, the secrecy outage probability of the random phase scheme is slightly better than that without IRS. Note that the performance of the random phase scheme does not change significantly as $N_s$ increases, because both Bob and Eve get similar gains from IRS.

\begin{figure}[htb]
\centering
\includegraphics[width = 0.45\textwidth]{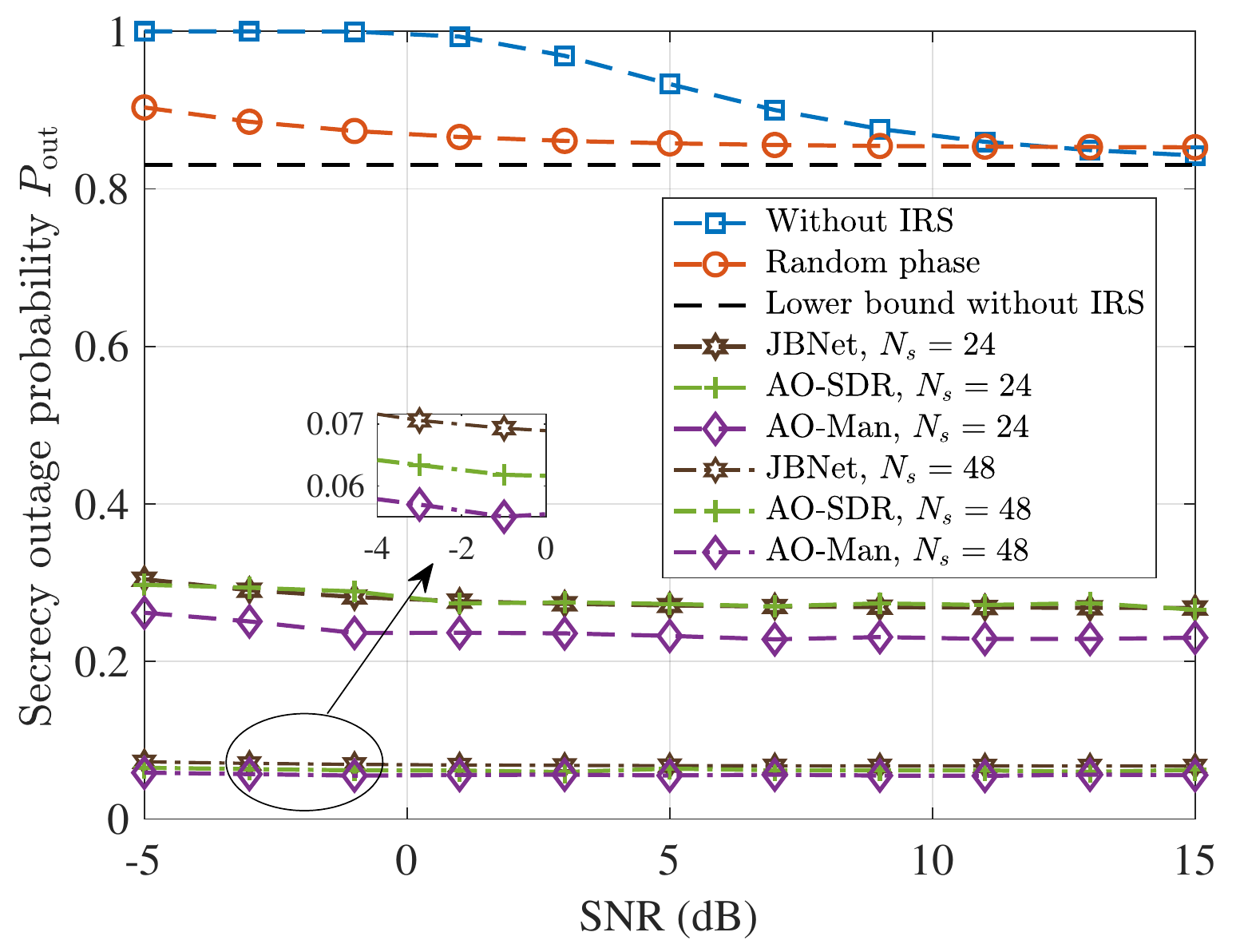}
\caption{Secrecy outage probability in terms of SNR when $R_s=3.5 ~\text{bit/s/Hz}$.}
\label{fig:5}
\end{figure}

Fig. \ref{fig:5} illustrates the impact of SNR on the secrecy outage probability. It is interesting to note that the performance of JBNet, AO-SDR algorithm, and AO-Man algorithm vary insignificantly in terms of SNR. However, this does not mean that increasing SNR is useless. As SNR increases, slightly increasing secrecy rate can be achieved with a low secrecy outage probability. Moreover, the performance of JBNet is very close to the two AO algorithms and much better than that without IRS when $N_s=48$. In particular, there is a lower bound without IRS when $P_t/\sigma_e^2 \rightarrow \infty$, where the proof is given in Appendix.  

\begin{figure}[htb]
\centering
\includegraphics[width = 0.45\textwidth]{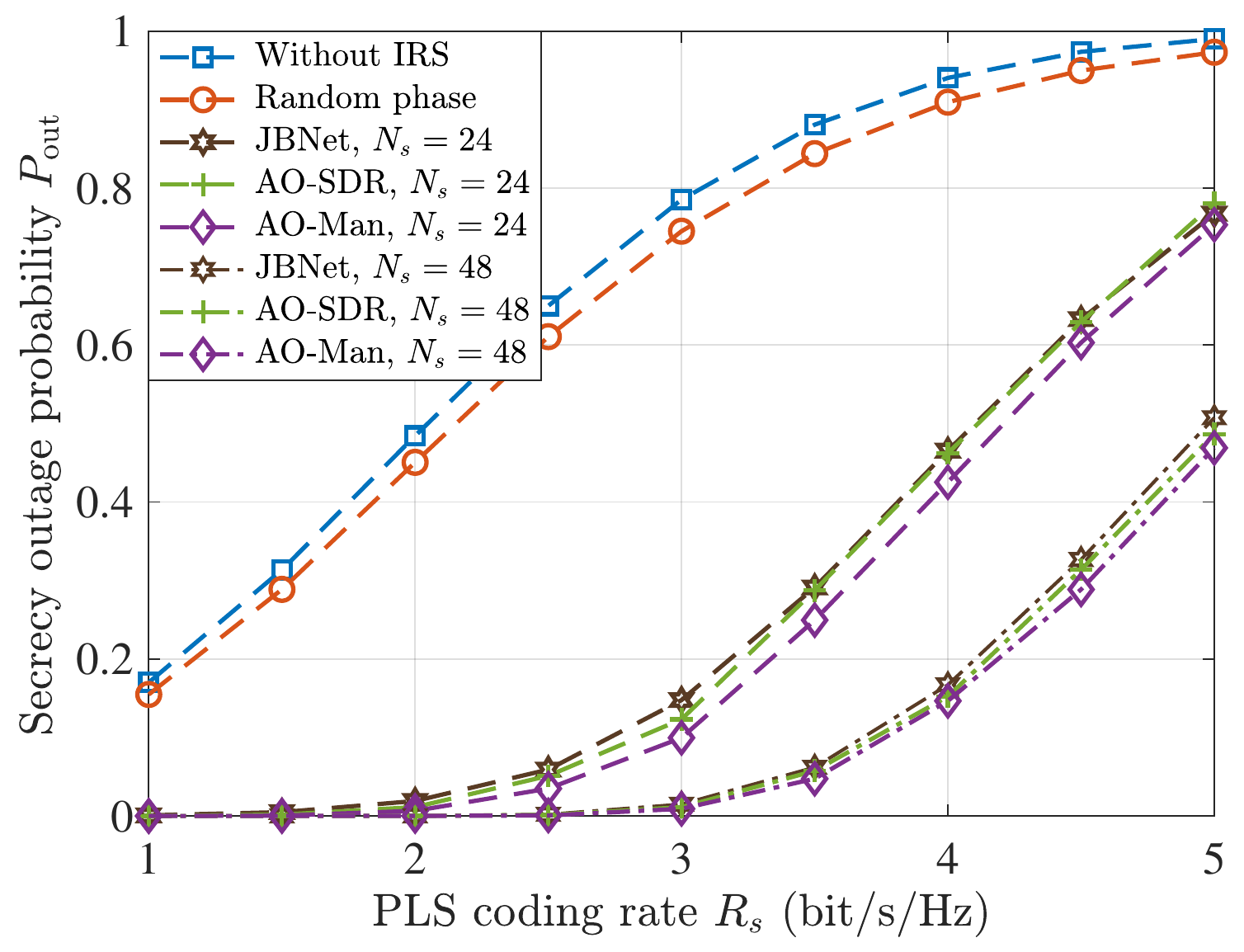}
\caption{Secrecy outage probability in terms of $R_s$ when $\text{SNR}=10~\text{dB}$.}
\label{fig:6}
\end{figure}

Fig. \ref{fig:6} depicts the impact of the PLS coding rate $R_s$ on the secrecy outage probability. As we can see from the figure, the secrecy outage probability of all schemes rises with an increasing $R_s$. In this case, some methods can be adopted to improve the PLS coding rate, such as increasing the number of IRS elements $N_s$, or increasing the numbers of antennas $N_t$ and $N_r$. For example, Fig. \ref{fig:6} also gives the comparison of simulation curves for $N_s=24$ and $N_s=48$.

\begin{figure}[htp]
\centering
\includegraphics[width = 0.45\textwidth]{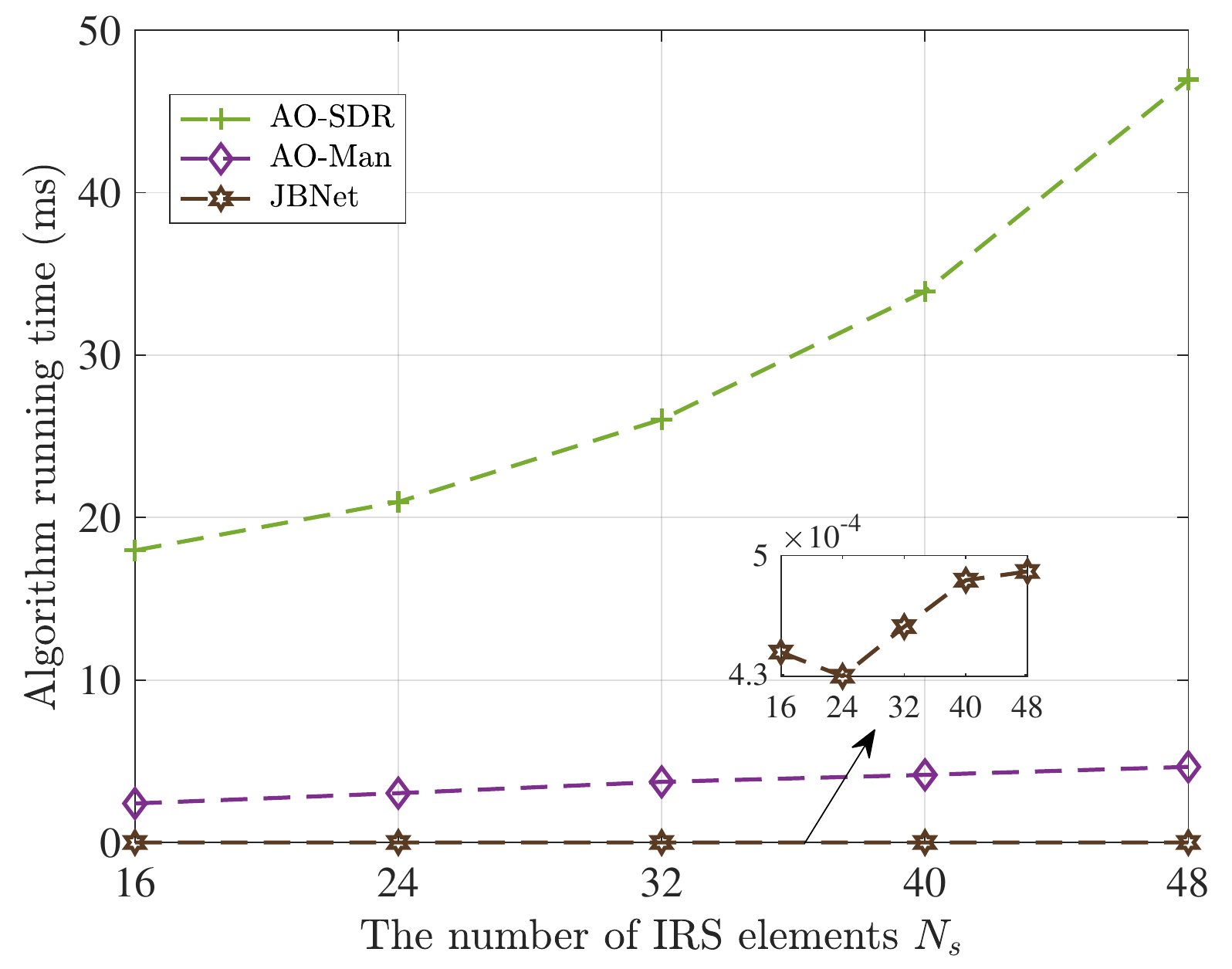}
\caption{Running time when $\text{SNR}=10~\text{dB}$ and $R_s=3.5 ~\text{bit/s/Hz}$.}
\label{fig:7}
\end{figure}

Fig. \ref{fig:7} shows the running time of total 20 simulations with different numbers of IRS elements $N_s$. For a fair comparison, all schemes are executed on the same hardware platform with Intel i5-8259U 2.3GHz CPU. It can be seen that the running time of the two AO algorithms increases significantly as $N_s$ increases, while the running time of JBNet remains almost unchanged. Note that $N_s=16$ has a longer running time than $N_s=24$, because $N_s = 16$ is the first round of the simulation and the program takes some extra time at startup. {According to~\cite{liu2022min}, the computational complexities of AO-SDR algorithm and AO-Man algorithm are $\mathcal{O}(K_{\max}(N_s^{4.5}+N_s^2N_t+N_t^3))$ and $\mathcal{O}(K_{\max}(N_s^{4}+N_s^2N_t+N_t^3))$, respectively, where $K_{\max}$ is the maximum number of iterations.  When the neural network has been trained, the parameters are fixed, so that solving the optimization problem Eq. (\ref{eq:9}) becomes some simple matrix computation, and the computational complexity of the DL-based approach is approximated as $\mathcal{O}(N_s^2+N_tN_rN_s)$\cite{elbir2021federated}. So, the proposed JBFNet has a lower computational complexity.}

\section{CONCLUSION}
We derived the expression of secrecy outage probability and proposed a low-complexity DL-based approach for joint beamforming design in an IRS-aided MIMOME system. To reduce the complexity in solving the optimization problem, the data pre-processing method, network structure, and loss function of the neural network have been designed. Simulation results showed that the secrecy performance of the proposed DL-based approach is close to that of the traditional AO algorithms, and the computational complexity is reduced significantly. In the future, we will extend the proposed model to more practical and challenging scenarios, such as IRS with a discrete phase shift, the amplitude and phase coupling, etc.

\appendix
In the case of PLS without IRS and ${P_t}/{\sigma_e^2} \rightarrow \infty$, $\phi$ in Eq. (\ref{eq:5a}) can be rewritten as 
\begin{subequations}
\label{eq:14}
\begin{align}
\phi &=\frac{\sigma_{e}^{2}\left(2^{C_{m}-R_{s}}-1\right)}{P_t} \\
&=\frac{\sigma_{e}^{2}\|\mathbf{H}_{b} \mathbf{b}\|^{2}}{\sigma^{2} 2^{R_{s}}}+\frac{\sigma_{e}^{2}}{P_t}\Big(\frac{1}{2^{R_{s}}}-1\Big) \leq \frac{\sigma_e^2 \lambda_{\max}}{\sigma^22^{R_s}},
\end{align}
\end{subequations}
where $\lambda_{\max}$ is the largest eigenvalue of $\mathbf{H}_b$. Therefore, the secrecy outage probability satisfies
\begin{subequations}
\label{eq:15}
\begin{align}
P_{\mathrm {out }}\left(R_{s}\right) &= \frac{1}{\Gamma\left(N_{e}\right)} \Gamma\left(N_{e}, \phi/\beta_d^2\right)\label{eq:16a}\\ 
&\geq \frac{1}{\Gamma\left(N_{e}\right)} \Gamma\Big(N_{e},  \frac{\sigma_e^2 \lambda_{\max}}{\sigma^22^{R_s}\beta_d^2}\Big)\label{eq:16b}.
\end{align}
\end{subequations}
Note that Eq. (\ref{eq:16b}) is a lower bound of the secrecy outage probability without IRS as shown in Fig. \ref{fig:5}.

\vfill
\end{document}